\documentclass{elsart}

\usepackage{epsfig}
\usepackage{amssymb}

\begin{document}

\begin{frontmatter}

\title{FLRW viscous cosmological models}

\author[label1]{G.S. Khadekar}\ead{gkhadekar@yahoo.com},
\author[label2]{Saibal Ray}\ead{saibal@associate.iucaa.in},
\author[label3]{X.-H. Meng}\ead{xhm@nankai.edu.cn}

\address[label1]{Department of Mathematics, R.T.M. Nagpur University, Mahatma Jyotiba Phule
Educational Campus, Amravati Road, Nagpur 440033, Maharastra, India}

\address[label2]{Department of Physics, Government College of Engineering and Ceramic Technology, 73 A.C.B. Lane, Kolkata
	700010, West Bengal, India}

\address[label3]{School of Physics, Nankai University, Tianjin 300071, P.R. China}

\begin{abstract}
In this paper we solve Friedmann equations by considering a universal media as a non-perfect fluid with bulk viscosity and is described by a general ``gamma law'' equation of state of the form  $p= (\gamma -1) \rho + \Lambda(t)$, where the adiabatic parameter $\gamma$ varies with scale factor $R$ of the metric and $\Lambda$ is the time dependent cosmological constant.  A unified description of the early evolution of the universe is presented by assuming the bulk viscosity and cosmological parameter in a  linear combination of two terms of the form: $\Lambda(t)=\Lambda_{0} + \Lambda_{1}\frac{\dot{R}}{R}$ and $\zeta = \zeta_{0} + \zeta_{1} \frac{\dot{R}}{R}$, where  $\Lambda_{0},\;\Lambda_{1},\; \zeta_{0}$ and  $ \zeta_{1}$ are constants,  in which an inflationary phase is followed by the radiation dominated phase. For this general gamma law equation of state, an entirely integrable dynamical equation to the scale factor $R$ is obtained along with its exact solutions.  In this framework we demonstrate that the model can be used to explain the dark energy dominant universe and for a special choice of the parameters we can explain the accelerating expansion of the universe also for two different phases, viz. combination of dark energy and dark matter phase as well as unified dark energy phase. A special physical check has been performed through sound speed constraint to validate the model. At last we obtain a scaling relation between the Hubble parameter with redshift. 
\end{abstract}

\begin{keyword}
FLRW cosmology; bulk viscosity; dark energy
\end{keyword}

\end{frontmatter}

\section{Introduction}
\label{intro}
Theoretical cosmology currently with several debating enters into a very interesting stage, which may imply the cosmic media or components and hence may not be completely possible to describe by the simplest perfect fluid model. Several cosmological observations on cosmic compositions and structure, from astrophysical as well as cosmological standpoints, indicate that our universe is undergoing a late time cosmic acceleration~[1-3]. In order to explain the accelerating expansion cosmologist have introduced a new  fluid which is known as {\it dark energy}. The first evidence in this direction, especially from the type Ia Supernovae~\cite{Riess2004,Jassal2005}  and WMAP satellite mission~\cite{Bennet2003},  is that we live in a favored spatially flat universe composed of approximately $4\%$   baryonic matter, $22\%$ dark matter and $74\%$ dark energy~\cite{Ren2006a} which is also confirmed by the Planck mission.  Some unified models have been proposed to detect the possibility of the unified assumption, like unified dark fluid model~[3,8-11] which assumes single fluid equation of state (EOS), inhomogeneous EOS and barotropic fluid dark energy, Chaplygin gas model, generalized Chaplygin gas etc.~[12-18]. 

The introduction of the viscosity into the cosmology has been investigated from different viewpoint~\cite{Padmanabhan1987,Gron1990}. In the following works~[19-37] the bulk viscosity in cosmology have extensively been studied in various aspects. Meng et al.~\cite{Meng2007} assumed the bulk viscosity as a linear combination of two terms: one is a constant and the other is proportional to the scalar expansion $\theta =3\frac{\dot R}{R}$ and discussed evolution of the universe through accelerating expansion and future singularity for FLRW model by using EOS of the form $p= (\gamma-1) \rho + p_{0}$, where $p_{0}$ is a parameter. Recently Khadekar et al.~\cite{Khadekar2015a} have solved the Friedmann equations with inhomogeneous EOS by considering bulk viscosity and time dependent parameter $\Lambda$~[1-3,40-48] as  linear combination of two terms in the forms: 
\begin{equation}
\label{eq1}\Lambda(t)=\Lambda_{0} + \Lambda_{1}\frac{\dot{R}}{R}, 
\end{equation} 

\begin{equation}
\label{eq2}\zeta(t) =\zeta_{0}+\zeta_{1}\frac{\dot{R}}{R},
\end{equation}
where $R$ is the scale factor and $\zeta_{0}$, $\zeta_{1}$, $\Lambda_{0}$,  $\Lambda_{1}$ all are constants.

It is to note in comparison to the abovementioned form of Meng et al.~\cite{Meng2007} that  here also one is a constant, however the other is proportional to Hubble parameter $H=\frac{\dot R}{R}$ and they~\cite{Khadekar2015a} successfully discussed the accelerating expansion of the universe evolution and future singularities in the framework of general theory of relativity. We would also like to mention that Khadekar and Ghogre~\cite{Khadekar2015b}  solved the Friedmann equations analytically as well as numerically in the framework of variable speed of light (VSL) theory by  assuming  $\Lambda(t)=\Lambda_{0} + \Lambda_{1}\frac{\dot{R}}{R}$ (exactly as Eq. (1)) and $\zeta(t)=\zeta_{0}+\zeta_{1}\frac{\dot{R}}{R}+\zeta_{2}\frac{\ddot{R}}{\dot R}$ (generalized form of Eq. (2)), where $\zeta_{2}$ is a constant, and treated the same issues of the accelerating expansion of the universe and future singularities. 

Madsen and Ellis~\cite{Madsen1998} presented the evolution of the universe for inflationary, radiation and matter dominated phase by considering ``gamma-law'' EOS $p = (\gamma-1) \rho$, where $\gamma$ is a function of the scale  factor $R$. Also Carvalho~\cite{Carvalho1996} has studied Robertson-Walker (RW) models in general relativity by assuming  that the adiabatic parameter $\gamma$ varies with cosmic time and presented unified description of early evolution of the universe  in which an inflationary phase is followed by the radiation dominated phase. Singh and his collaborators~\cite{Singh2006a,Singh2006b,Singh2007a,Singh2007b,Singh2012} discussed FLRW model by using the variable adiabatic parameter $\gamma$ as a function $R$ and also studied the evolution of the universe as it goes from an inflationary  phase to a radiation dominated phase.  All these treatments therefore suggest that instead of a pure constant one can assume $\gamma$ as a variable throughout the history of the universe. 

Under the above motivating background the present study is based on the time dependent viscosity, inhomogeneous EOS and Hubble parameter dependent EOS as considered in their work by Nojiri and Odintsov~\cite{Nojiri2005a}. We shall present analytically the equation of the scale factor derived from the inhomogeneous EOS given by~\cite{Khadekar2015a}  and study the unified description of the early evolution of the universe by using the EOS  parameter $\gamma$ which varies with the scale factor $R$ proposed by Carvalho \cite{Carvalho1996}, where an inflationary phase is followed by the radiation dominated phase. 

This paper is organized as follows: in Sec. 2 we describe our model and provide exact solutions of the scale factor for two different time dimensions: $T_ {2} \rightarrow \infty$ and  $T_ {1} \rightarrow \infty$.  We study the evolutions of the universe as it goes from an inflationary phase to a radiation dominated era by using general gamma law EOS parameter $\gamma$ depending on the scale factor $R$. In Sec. 3 we propose  two consequences of the model, viz. (i) combination of dark energy and dark matter, and (ii) unified dark energy. A particular solution for $a \rightarrow 0$ is obtained under this case of unified dark energy whereas in Sec. 4 a discussion on the sound speed constraint has been done as a tool to verify the acceptibility of the proposed model. Finally, in Sec. 5 we present our discussion and conclusions. However, we also provide an interpretation of the case  $T_ {1} \rightarrow \infty$ for both the phases in {\bf Appendix}.

\section{The field equations and general results of the model }
We consider the FLRW metric of the form
\begin{equation}
\label{eq3}  ds^{2}= -dt^{2}+R^{2}(t)(dr^{2}+r^{2} d\theta^{2} + r^{2} sin^{2}\theta d\phi^{2}),  
\end{equation}
where $R$ is the scalar factor. 

The Einstein field equations take the usual form 
\begin{equation}
\label{eq4} R_{\mu\nu}-\frac{1}{2}g_{\mu\nu}R = 8 \pi GT_{\mu\nu}.
\end{equation}

In the FLRW cosmology with bulk viscosity the stress energy momentum tensor can be written as 
\begin{equation}
\label{eq5} T_{\mu\nu} = (p+\rho) U_{\mu} U_{\nu} + p g_{\mu\nu}-\zeta\theta H_{\mu\nu},         
\end{equation}
where $\zeta$ is the bulk viscosity, $\theta$ the expansion factor defined by $\theta = 3\dot{R}/R$ and the projection tensor $H_{\mu\nu}$ is defined by $H_{\mu\nu} = g_{\mu\nu} + U_{\mu} U_{\nu}$ with $U_{\mu\nu}$ being the four velocity and the fluid on the comoving coordinates  are represented by $p$ and $\rho$,  respectively the pressure and density. 

For the FLRW model (1) the Einstein field equations are given by 
\begin{equation}
\label{eq6} H^2= \frac{8 \pi G}{3}\rho,
\end{equation} 
\begin{equation}
\label{eq7} \dot H +H^2= - \frac{4 \pi G}{3}(\rho + 3\bar{p}),
\end{equation}
where $\bar{p}$ is an equivalent pressure defined by $\bar{p}= p-\zeta \theta$ and dot ${\bf(\; \bf \dot{} \; )}$ stands for  differentiation with respect to time.

The conservation equation for energy $T^{0 \nu}_{;0}=0$ yields
\begin{equation}
\label{eq8} \dot \rho +(\rho +\bar{p})\theta = 0.
\end{equation}

We consider the inhomogeneous EOS of the form \cite{Khadekar2015a}
\begin{equation}
\label{eq9}p= (\gamma-1)\rho +\Lambda(t), 
\end{equation}
where $\Lambda(t)$ is the time dependent parameter, the erstwhile Einstein cosmological constant.

After eliminating $p$ and $\rho$ from Eq. (8) with the help of Eq. (1), Eq. (2) and Eq. (9), we get
\begin{equation}
\label{eq10} \dot H = - \frac{3\bar{\gamma}}{2} H^2+ 4 \pi G (3\zeta_{0} - \Lambda_{1}) H - 4 \pi G \Lambda_{0} ,
\end{equation}
where $\bar{\gamma} = \gamma - 8 \pi G \zeta_{1}$. One can observe that in the EOS  the equivalent effect of the second term in $\zeta$ is to change the parameter  $\gamma$ to  $\bar{\gamma}$.

Now we define \cite{Khadekar2015a}
\begin{equation}
\label{eq12} 4 \pi G (3\zeta_{0} - \Lambda_{1})=\frac{1}{T_{1}},
\end{equation}
\begin{equation}
\label{eq13} -4 \pi G \Lambda_{0}=\frac{1}{T_{2}^{2}}.
\end{equation}
Note that the dimension of the above two terms  $4 \pi G (3\zeta_{0} - \Lambda_{1})$ and $-(4 \pi G \Lambda_{0})$ is $time^{-1}$ and $time^{-2}$ respectively.

Then from  Eq. (9) the dynamical equation of the scale factor $R(t)$ can be written as 
\begin{equation}
\label{eq14} \dot H = -\frac{3\bar{\gamma}}{2}H^2+\frac{1}{T_{1}} H + \frac{1}{T_{2}^{2}}.
\end{equation}

From the above Eqs. (10)-(12), it can be observed that the five parameters $\gamma, \; \zeta_{0},\; \zeta_{1},\; \Lambda_{0},$ and $ \Lambda_{1}$ are condensed to three parameters $\bar \gamma,\; T_{1}$, and $T_{2}$ in a single Eq. (13). It is also to note that if $T_{1} \rightarrow \infty$  then $3\zeta_{0} = \Lambda_{1}$ and if $T_{2} \rightarrow \infty$ then the effect of $\Lambda_{0}$ can be neglected. 

The Eq. (13) can also be written in a more convenient way as
\begin{equation}
\label{eq15}  H' = -\frac{3\bar{\gamma}}{2}\frac{H}{R}+\frac{1}{T_{1}}\frac{1}{ R}  + \frac{1}{T_{2}^{2}} \frac{1}{HR},
\end{equation}
where a prime $(')$ denotes differentiation with respect to the scale factor $R$.

In the following we solve Eq. (14) for the cases: (i) $T_{2} \rightarrow \infty$  and (ii)  $T_{1} \rightarrow \infty$. However, we have given here detail treatment only of the former one whereas the latter case can be obtained in the {\bf Appendix} with proper interpretation.

\subsection{When $T_{2} \rightarrow \infty$ }
In this case Eq. (14) can be expressed as 
\begin{equation}
\label{eq16}  H' + \frac{3\bar{\gamma}}{2}\frac{H}{R}= \frac{1}{T_{1}}\frac{1}{ R}.
\end{equation}

We  obtain the first integral of the  above equation by using the functional form of  $\gamma$  which depends on the scale factor as proposed by Carvalho~\cite{Carvalho1996}  in the  following structure 
\begin{equation}
\label{eq17} \gamma
(R)=\frac{4}{3}\frac{A(R/R_{*})^{2}+(\frac{a}{2})
(R/R_{*})^{a}}{A(R/R_{*})^{2}+(R/R_{*})^{a}},
\end{equation}
where $A$ is a constant and $a$ is the free parameter related to the
power of cosmic time and lies $0\leq a <1$. Here $R_{*}$ is
certain reference value such that if $R\ll R_{*}$, inflationary
phase of the evolution of the universe is obtained ($\gamma\leq
\frac{2a}{3} $) and for $R\gg R_{*}$, we have a
radiation dominated phase ($\gamma=\frac{4}{3}$).

After substituting above value of $\gamma$ in Eq. (15), we get
\begin{eqnarray}
\label{eq18} \frac{H}{R^{\alpha_{0}}}[A(R/R_{*})^{2}+(R/R_{*})^{a}]= \nonumber \\ 
\frac{1}{T_{1}}\int \frac{1}{ R^{\alpha_{0}}R}[A(R/R_{*})^{2}+(R/R_{*})^{a}]dR +C_{0},
\end{eqnarray}
where $C_{0}$ is a constant of integration and $\alpha_{0}= 12\pi G \zeta_{1}.$

Now we solve Eq. (17) for two different phases of the universe, viz. {\it inflationary} and {\it radiation} dominated phases respectively. 

\subsubsection{Inflationary phase:} For this phase $ R<<R_{*}$ and the above Eq. (17) reduces to 
\begin{equation}
\label{eq19} H= \frac{1}{\beta_{0} T_{1}}+ \frac{C_{0} R^{\alpha_{0}}}{(R/R_{*})^{a}},
\end{equation}
where $\beta_{0}=(a-\alpha_{0})$.

If $C_{0}=0$ in Eq. (18) then $H$ has constant value in both the phases and therefore we have the exponential expansion. However, when $C_{0}\ne 0$ then with the initial condition $R(t_{0}) = R_{0}$ and $H(t_{0}) = H_{0}$, we get
\begin{equation}
\label{eq20}  C_{0} = \frac{R^{\beta_{0}}}{R_{*}}\left(H_{0} -\frac{1}{\beta_{0}T_{1}}\right).
\end{equation}

After inserting this value of $C_0$ in Eq. (18), we get
\begin{equation}
\label{eq21}  H = (R_{0}/R)^{\beta_{0}}\left(H_{0} -\frac{1}{\beta_{0} T_{1}}\right) + \frac{1}{\beta_{0} T_{1}}.
\end{equation}

From Eq. (20) the scale factor and energy density can respectively be obtained as 
\begin{eqnarray}
\label{eq22} R= R_{0}\left[ \left( 1+ \frac{1}{\beta_{0} T_{1}}[H_{0} -\frac{1}{\beta_{0} T_{1}}]\right)e^{(t-t_{0})/T_{1}} -  \frac{1}{\beta_{0} T_{1}}[H_{0} -\frac{1}{\beta_{0} T_{1}}] \right]^{1/\beta_{0}},
\end{eqnarray}

\begin{equation}
\label{eq23}
\rho= \frac{3}{8\pi G}\left[\frac{\left( 1+ \frac{1}{\beta_{0} T_{1}}[H_{0} -\frac{1}{\beta_{0} T_{1}}]\right)e^{(t-t_{0})/T_{1}}-[H_{0} -\frac{1}{\beta_{0} T_{1}}][\frac{1}{\beta_{0} T_{1}}-T_{1}\beta_{0}]}{ T_{1}\beta_{0}\left( 1+ \frac{1}{\beta_{0} T_{1}}[H_{0} -\frac{1}{\beta_{0} T_{1}}]\right)e^{(t-t_{0})/T_{1}}-[H_{0} -\frac{1}{\beta_{0} T_{1}}]} \right]^{2}.
\end{equation}

The above equation is valid for $\bar\gamma \ne 0$. For $\bar \gamma=0$, Eq. (15) reduces to
\begin{equation}
\label{eq24} H= H_{0}+ \frac{1}{T_{1}}ln(R/R_{0}).
\end{equation}

The scale factor and energy density can respectively be obtained as 
\begin{equation}
\label{eq25} R= R_{0}exp\left[ T_{1} H_{0}(e^{(t-t_{0})/T_{1}}-1)\right],
\end{equation}

\begin{equation}
\label{eq26} \rho =\frac{3}{8\pi G} H^2_{0} e^{2(t-t_{0})/T_{1}}.
\end{equation}

It is observed that the solution of the scale factor for $\bar \gamma=0$ does not possess any future singularity, the so-called {\it big rip}. 

\subsubsection{Radiation phase:} For this phase $ R>>R_{*}$ and we have 
\begin{equation}
\label{eq27}  H = (R_{0}/R)^{x_{0}}\left(H_{0} -\frac{1}{x_{0} T_{1}}\right) + \frac{1}{x_{0} T_{1}},
\end{equation}
where $x_{0}=(2-\alpha_{0})$.

From Eq. (26) the scale factor and energy density can respectively be obtained as 
\begin{eqnarray}
\label{eq28} R= R_{0}\left[ \left( 1+ \frac{1}{x_{0} T_{1}}[H_{0} -\frac{1}{x_{0} T_{1}}]\right)e^{(t-t_{0})/T_{1}} -  \frac{1}{x_{0} T_{1}}[H_{0} -\frac{1}{x_{0} T_{1}}] \right]^{1/x_{0}},
\end{eqnarray}

\begin{equation}
\label{eq29}\rho= \frac{3}{8\pi G}X _{1}^2,
\end{equation}
where 
\begin{equation}
X_{1}= \left[\frac{\left( 1+ \frac{1}{x_{0} T_{1}}[H_{0} -\frac{1}{x_{0} T_{1}}]\right)e^{(t-t_{0})/T_{1}}-[H_{0} -\frac{1}{x_{0} T_{1}}][\frac{1}{x_{0} T_{1}}-T_{1}x_{0}]}{ T_{1}x_{0}\left( 1+ \frac{1}{x_{0} T_{1}}[H_{0} -\frac{1}{x_{0} T_{1}}]\right)e^{(t-t_{0})/T_{1}}-[H_{0} -\frac{1}{x_{0} T_{1}}]} \right]^{2}.\nonumber
\end{equation}

The above equation is valid for $\bar\gamma \ne 0$.

To illustrate the parameters involved in the above solution set more clearly, we draw some graphics in Figs. 1 and 2. For this we set $\frac{3}{8 \pi G}=1,\; H_{0}=1,\; \beta_{0}=1$ and the values of the other parameters are given in the legend and the caption of each figures. Fig. 1 shows the evolution scenario of the scale factor.  Similarly  Fig. 2 dictates about the relationship between density $\rho$ and  $(t- t_{0})$  for the inflationary phase.


\begin{figure*}[thbp]
\centering
\includegraphics[width=0.45\textwidth]{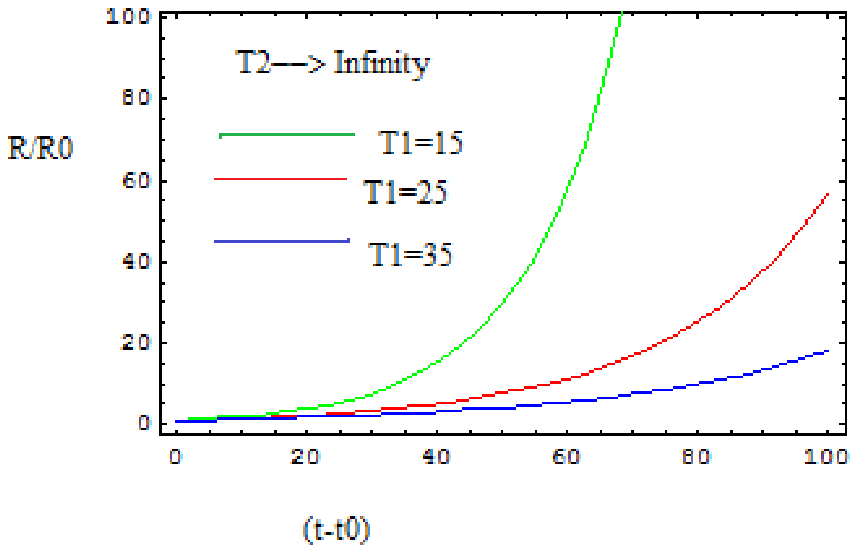}
\includegraphics[width=0.45\textwidth]{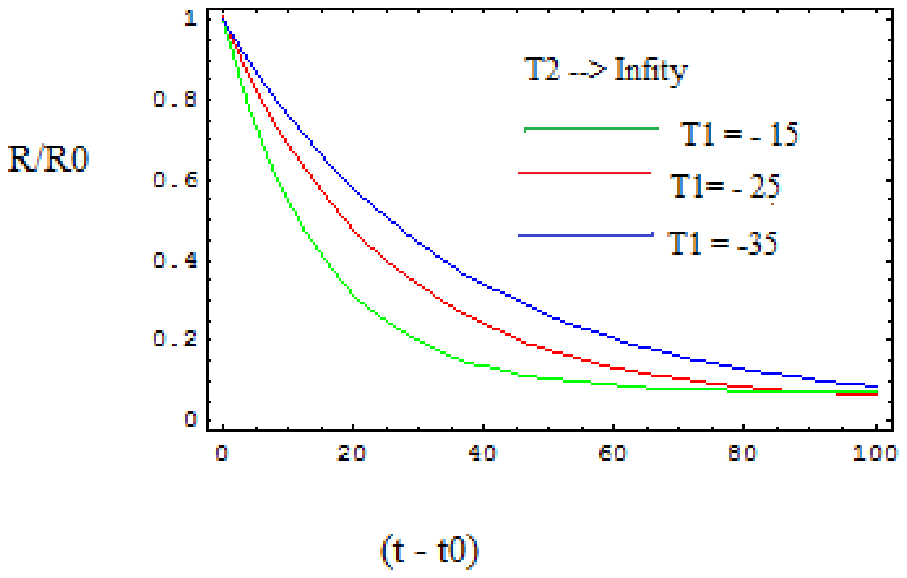}
\caption{ Relation between $R/R_{0}$ and  $t-t_{0}$ with the values of $T_{1}=15,\; 25,\; 35$ (left panel) and $T_{1}=-15,\; -25,\; -35$ (right panel) 
when $T_{2} \rightarrow \infty$ for the inflationary phase $R << R_{*}$}
\end{figure*}



\begin{figure*}[thbp]
\centering
\includegraphics[width=0.45\textwidth]{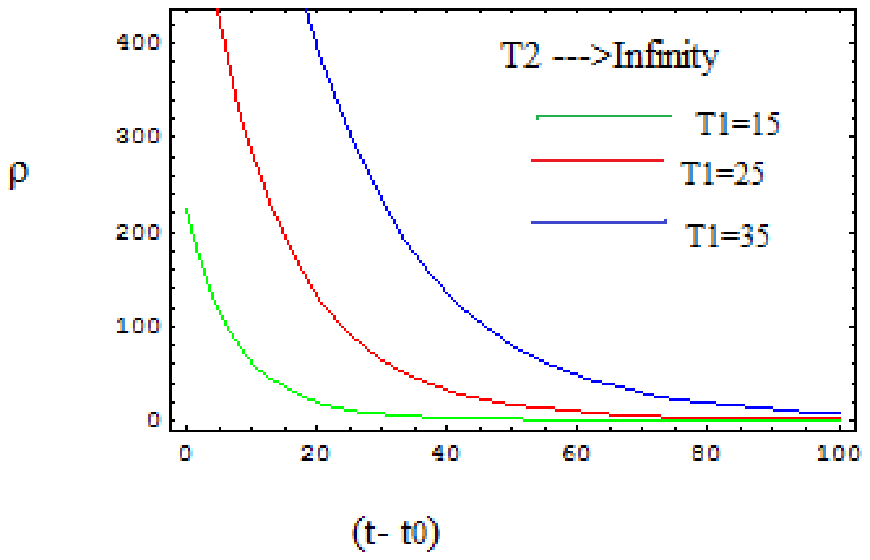}
\includegraphics[width=0.45\textwidth]{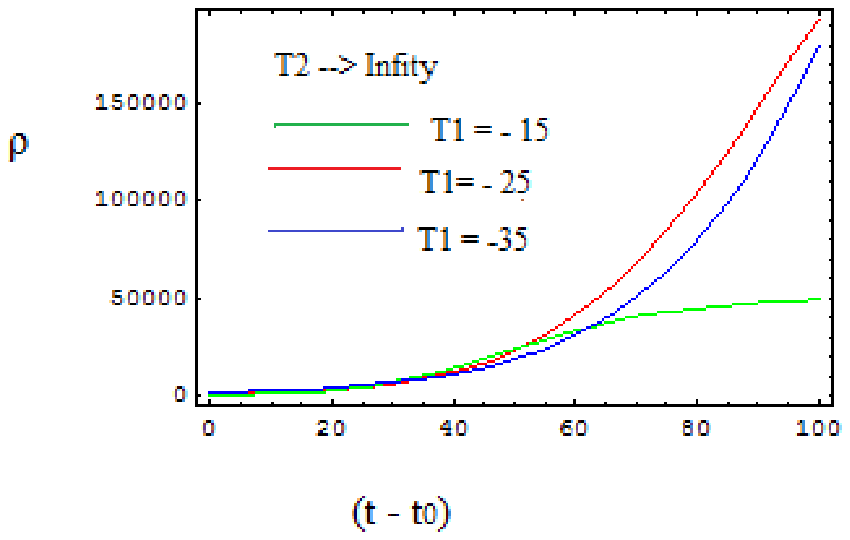}
\caption{Relation between density $\rho$ and  $t-t_{0}$ with the values of $T_{1}=15,\; 25,\; 35$  (left panel) and $T_{1}=-15,\; -25,\; -35$ (right panel)
when $T_{2} \rightarrow \infty$ for the inflationary phase $R << R_{*}$}
\end{figure*}


\section{Physical consequences of the model}
\subsection{Combination of dark energy and dark matter}
In the $\Lambda$-CDM model~\cite{Ren2006a}
\begin{equation}
\label{eq30} H^2 = H_{0}^2[\Omega_{m}(1+z)^3 +\Omega_{\Lambda}],
\end{equation}
where $H_{0}$ is the Hubble parameter in the present epoch, $z= (R_{0}/R -1)$ is the redshift, $\Omega_{m}$ and $\Omega_{\Lambda}$ 
are the cosmological density parameter of matter and dark energy respectively. 

Now in our case the solution of Eq. (14) is in the form of Eq. (20) 
when $ T_{2}\rightarrow \infty $ and in terms of $H(z)$ for the inflationary phase $(R << R_{*})$
\begin{equation}
\label{eq31}  H(z) = \left(H_{0} -\frac{1}{\beta_{0} T_{1}}\right) (1+z)^{\beta_{0}} + \frac{1}{\beta_{0} T_{1}}.
\end{equation}

The $H_{d}(z)$ for the dark energy is 
\begin{equation}
\label{eq32}  H_{d}(z) = H_{0}[\bar \Omega (1+z)^{\beta_{0}} + (1-\bar \Omega)],
\end{equation}
where $\bar \Omega= (1-\frac{1}{\beta_{0}H_{0}T_{1}}).$

It should be noted that Eq. (31) is valid for $\bar \gamma \ne 0$, and for $\bar\gamma=0$ by directly solving Eq. (15) one can get 
\begin{equation}
\label{eq33}  H_{d}(z) = H_{0}\left[1-\frac{1}{H_{0} T_{1}} ln(1+z)\right].
\end{equation}

Hence the {\it H-z} relation can be obtained as 
\begin{equation}
\label{eq34}  H^2 = H_{0}^2 \left\{ \Omega_{m} [\bar \Omega (1+z)^{\beta_{0}} + (1-\bar \Omega)]^2 + (1-\Omega_{m})\right\}.
 \end{equation}

Thus we can recover an analogous result obtained earlier by Ren and Meng~\cite{Ren2006a} in a different context.

On the other hand, for the radiation dominated phase the solution of Eq. (14) is in the form of (26) when $ T_{2}\rightarrow \infty $ and in terms of $H(z)$ is given by 
\begin{equation}
\label{eq35}  H(z) = \left(H_{0} -\frac{1}{x_{0} T_{1}}\right) (1+z)^{x_{0}} + \frac{1}{x_{0} T_{1}}.
\end{equation}

The $H_{d}(z)$ for the dark energy is 
\begin{equation}
\label{eq36}  H_{d}(z) = H_{0}[\bar \Omega_{1} (1+z)^{x_{0}} + (1-\bar \Omega_{1})],
\end{equation}
where $\bar \Omega_{1}= (1-\frac{1}{x_{0}H_{0}T_{1}}).$

Note that Eq. (36) is valid for $\bar \gamma \ne 0$. 

The {\it H-z} relation is 
\begin{equation}
\label{eq37}  H^2 = H_{0}^2 \left\{ \Omega_{m} [\bar \Omega_{1} (1+z)^{x_{0}} + (1-\bar \Omega_{1})]^2 + (1-\Omega_{m})\right\}.
 \end{equation}

\subsection{Unified dark energy} 
The Eq. (13) can be written as 
\begin{equation}
\label{eq38} \frac{\ddot{R}}{R} = -\frac{(3\bar{\gamma}-2)}{2}\frac{\dot{R}^2}{R^2}+\frac{1}{T_{1}}\frac{\dot{R}}{R} + \frac{1}{T_{2}^{2}}.
\end{equation}

We observe that the three terms on the right hand side of Eq. (38) are proportional to ($\frac{\dot{R}}{R})^2,\; (\frac{\dot{R}}{R})^{1}$ and ($\frac{\dot{R}}{R})^{0}$ respectively. Therefore in the following we shall separately study the effect of the above three terms. If the first term be dominant than the other two terms then from Eq. (38) we get 
\begin{equation}
\label{eq39} \frac{\ddot R}{ R}= -\frac{(3 \bar \gamma-2)}{2}\frac{\dot R^2}{R^2}\Rightarrow \frac{dH}{dR}+ \frac{3\bar{\gamma}}{2R}H=0.
\end{equation}

After integrating the above equation one can obtain
\begin{equation}
\label{eq40} H(R)= C_{1}exp\left( \frac{-3}{2}\int \frac{\bar\gamma(R)}{R}dR \right),
\end{equation}
where $C_{1}$ is the constant of integration.

By using the expression of $\gamma(R)$ from Eq. (16) we get 
\begin{equation}
\label{eq41} H(R)= \frac{C_{1} R^{(3\alpha_{0}/2)}}{[A(R/R_{*})^{2}+(R/R_{*})^{a}]^{1/4}}.
\end{equation}

If $H=H_{*}$ for $R=R_{*}$, then we have the relation between $A$ and $C_{1}$ as 
\begin{equation}
\label{eq42} C_{1}= \frac{H_{*}(A +1)^{1/4}}{R^{(3\alpha_{0}/2)}}.
\end{equation}

For  the inflationary phase $(R<<R_{*})$, Eq. (41) gives the scale factor $R$ as 
\begin{equation}
\label{eq43} R^{(a - 6\alpha_{0})/4} =  \frac{1}{4}C_{1} R_{*}^{a/4} \left(a - 6\alpha_{0}\right)t, 
\end{equation}
and for radiation dominated phase $(R>>R_{*})$
\begin{equation}
\label{eq44} R^{(1-3\alpha_{0})/2} =  \left(\frac{2C_{1} R_{*}}{(1-3\alpha_{0}) A^{1/4}}\right)t.
\end{equation}

If the second term is dominant then from Eq. (38) we get
\begin{equation}
\label{eq45} \frac{\ddot{R}}{R} = \frac{1}{T_{1}}\frac{\dot{R}}{R},
\end{equation}
which essentially describes the effective viscosity~\cite{Wang2014}.

The solution of the above equation is already discussed by Ren and Meng~\cite{Ren2006a}  in the following form
\begin{equation}
\label{eq46} H(z)= (H_{0} -\frac{1}{T_{1}})(z+1) + \frac{1}{T_{1}}.
\end{equation}

In terms of the scale factor $R$ we get
\begin{equation}
\label{eq47}R=  R_{0}[ H_{0}( e^{(t -t_{0})/T_{1}}-1)-1].
\end{equation}

\subsubsection{A particular solution of the model }
In this case we specially study the solution of the unified dark energy model in the limit $a \rightarrow 0$. Then Eq. (41) readily provides
\begin{equation}
\label{eq48} H(R)= \frac{C_{1} R^{(3\alpha_{0}/2)}}{[A(R/R_{*})^{2}+1]^{1/4}}.
\end{equation}

After integration we get 
\begin{equation}
\label{eq49} C_{1} t= \int \frac{[A(R/R_{*})^{2}+1]^{1/4}}{ R^{(3\alpha_{0}/2)+1}}dR.
\end{equation}

Again in the limit of very small $R$ for $(R<< R_{*})$ the second term becomes dominant and it gives
\begin{equation}
\label{eq50} C_{1} t= \int \frac{dR}{ R^{(3\alpha_{0}/2)+1}}.
\end{equation}

After solving above integration we get 
\begin{equation}
\label{eq51} R=  \left[\frac{-3\alpha_{0}C_{1}}{2}\right]^{-2/3\alpha_{0}} t^{-(2/3\alpha_{0})}.
\end{equation}

From this equation it is observed that for $\alpha_{0} >0,\; C_{1}>0$, which leads to contraction. As $t\rightarrow -\infty$, we find that $R\rightarrow 0$ and 
the model start from infinite past with zero power volume. Thus for $a=0$ the universe is infinitely old and we have inverse power law. 

On the other hand, in the the limit $a \rightarrow 0$ for the radiation dominated phase  $(R>> R_{*})$, the first term in the denominator of Eq. (41) dominates and then after simplifying we get
\begin{equation}
\label{eq52}    R =  \left[\frac{ R^{1/2}_{*}C_{1}}{2(1 -3\alpha_{0}) A^{1/4}}\right]^{2/(1-3\alpha_{0})} t^{2/(1-3\alpha_{0})}.
\end{equation}

Here the scale factor has the form $R \propto t^{2/(1-3\alpha_{0})}$, which shows the power law expansion of the universe.

From Eq. (41), a unified expression for the deceleration parameter $q$ can be given in terms of the scale factor as 
\begin{eqnarray}
\label{eq53}    q = -\left(1+ \frac{R}{H}\frac{dH}{dR}\right)= -\left[\frac{ (3\alpha_{0}/2) A(R/R_{*})^{2} + (3\alpha_{0}/2 +1)}{A ( R/R_{*})^{2}+1}\right].
\end{eqnarray}

From the above expression of deceleration parameter therefore we can observe that $q$ varies from $q=-(\frac{3\alpha_{0}}{2} +1)$ for inflationary phase ($ R<<R_{*}$) to $q= -\frac{3\alpha_{0}}{2}$ for radiation dominated phase ($ R>>R_{*}$). The deceleration parameter $q$ is positive for $\alpha_{0}<-2/3$  and negative for $\alpha_{0}>-1/6$. Similarly for the radiation dominated phase the universe decelerates for    $\alpha_{0}>0$ and accelerates for $\alpha_{0}<0$. Thus, there is a clear signature of flip-flop in our evolutionary scenario as can be observed in the models provided by Usmani et al.~\cite{Usmani2008} and Yadav et al.~\cite{Yadav2016} in different context.

\section{Sound speed constraint}
It is essential that in an acceptable model the sound speed ($c_{s}^{2}=\frac{\partial p}{\partial \rho}$) should be constant and in the range [0, 1]~\cite{Xia2008,Xu2012,Avelino2015} . 

Now, using Eq. (9) the EOS between $p$ and $\rho$ is given by 
\begin{equation}
\label{eq55} p = (\bar\gamma-1)\rho-\frac{2\sqrt{\rho}}{\sqrt{3}\kappa T_{1}}-\frac{2}{T_{2}^{2}\kappa^{2}}, 
\end{equation}
where $\kappa =8\pi G$.


\begin{figure*}[thbp]
\centering
\includegraphics[width=0.4\textwidth]{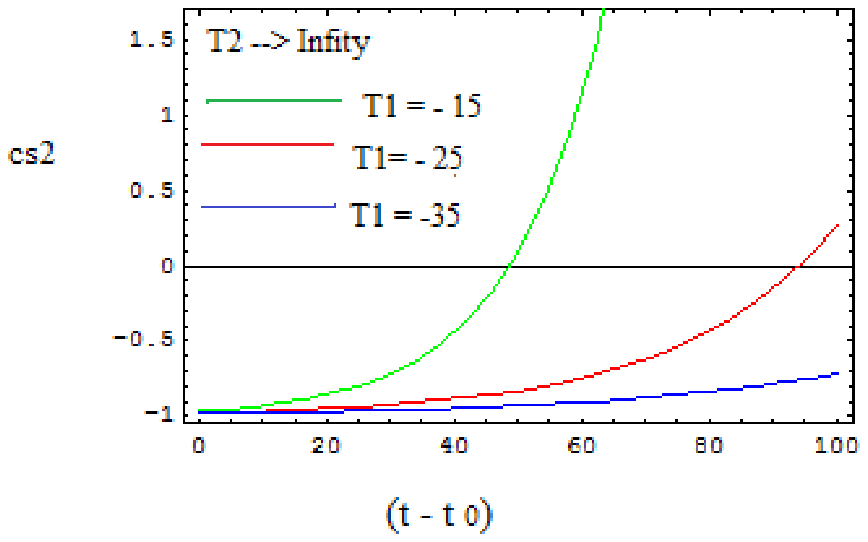}
\includegraphics[width=0.5\textwidth]{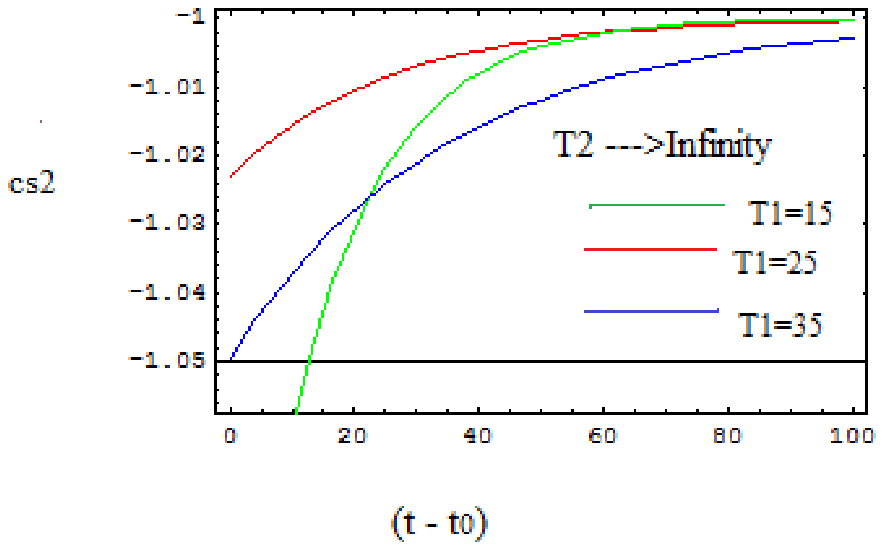}
\caption{Relation between the  square of the sound speed $c_{s}^2$  and time $t-t_{0}$ with the values of $T_{1}=-15,\; -25,\; -35$ (left panel) 
and  $T_{1}=15,\; 25,\; 35$ (right panel)}
\end{figure*}


Therefore, the explicit form of the sound speed in terms of our model parameters can be written as
\begin{equation}
\label{eq56} c_{s}^{2}=\frac{\partial p}{\partial \rho}= \bar\gamma-1- \frac{1}{\sqrt{3}\kappa T_{1} \sqrt{\rho}}.
\end{equation}
In our case, if $\bar \gamma$ is around zero,  then from Eq. (25) we can see that $\rho$ approaches to constant in the early time. i.e. 
as $(t-t_{0}) \rightarrow 0$, $\rho \rightarrow \frac{3H_{0}^2}{\kappa}$. Hence we can see that in the present model the sound speed approaches to constant in the early times. The feature is shown in Fig. 3.

\section{Discussions and Conclusions}
Now-a-days cosmology has really entered in a golden age as lots of observational data-sets are coming up to clarify various cosmic phenomena. In this paper, we have considered the approach of unified EOS which has been enabled us to describe the universe contents to several 
 fundamental issues in cosmological evolution from unified viewpoint. We have found the analytical solutions for the scale factor $R$, 
the Hubble parameter $H$, the energy density of matter $\rho$, the deceleration parameter $q$ by using the EOS of the form (9)
 with the assumption that $\Lambda$ and $\zeta$ are in the form of  Eqs. (1) and (2) respectively.  For this we have taken the
 adiabatic parameter $\gamma$ as a function of the scale factor $R$ and  obtained the cosmological solution for the two earlier
 phases of the evolutions of the  universe for $T_{2} \rightarrow \infty$  and $T_{1} \rightarrow \infty$   and have discussed the cases
 for the values of $\bar\gamma \ne 0$ and $\bar\gamma = 0$.

Astrophysical and cosmological observations are usually interpreted in terms of dark matter and dark energy~\cite{Arbey2005}. Two
 interpretations of the present model are proposed which are as follows:  (i) the EOS describes the dark energy combined  with dark matter in the
 universe media or EOS describes the dark matter with viscosity, which is mixed with dark energy for $\Lambda$ term of Hubble
 parameter $H$, and (ii) the EOS describes the dark matter and the dark energy, so there is a single fluid to show functions in the
 universe. In this case we prefer to the choice of the parameter $\bar \gamma \approx 0, \;\; T_{1} < 0$ and  $T_{2}$  is a real
 number or $T_{2}^{2} >0.$ 

It is already mentioned that in Eq. (38), the term $\frac{1}{T_{1}}\frac{\dot R}{R}$ describes the effective viscosity~\cite{Wang2014}. From the physical point of view Eq. (38) naturally contains the dissipative process in the cosmological evolution. 
It is observed that each term in the right hand side of Eq. (38) accounts for the time dependent bulk viscosity or the variable cosmological constant. 

We have shown that for the particular solutions in the limit $a \rightarrow 0,$ the universe is infinitely old, since $R\rightarrow 0$ as  $t \rightarrow -\infty,$  for $R<<R_{*}$  whereas for $R >> R_{*}$ the scale factor has the form $R \propto t^{2/(1-3\alpha_{0})}$. This evidently shows the power law expansion of the universe. We observe that the deceleration parameter $q$ varies from $q= -(\frac{3\alpha_{0}}{2}+ 1)$ for
 inflationary phase to $q= -\frac{3\alpha_{0}}{2}$ for radiation dominated phase and receives a smooth turn overn via the factor `-1'. Similarly for the radiation dominated phase the
 universe is decelerates for $\alpha_{0} >0$ and accelerates for $\alpha_{0} <0$ which indicates a flip-flop in the evolutionary scenario.

If we define the effective EOS as usual $\omega= p/\rho$ then Eq. (55) yields 
\begin{equation}
\label{eq57}  \omega = \frac{p}{\rho} = (\bar \gamma-1)-\frac{2}{\sqrt{3}\kappa T_{1}\sqrt{\rho}}-\frac{2}{T_{2}^{2}\kappa^{2} \rho}.
\end{equation}

The observational constraints indicates that the value of the EOS parameter $\omega=p/\rho$ is around $-1$. It is quite probably 
$\omega<-1$, known as phantom region and even more mysterious in the cosmological evolution~\cite{Riess2004,Jassal2005}. In our case for 
$\bar \gamma=0$ and there is no $\omega = -1$ crossing. However, if we consider ($\frac{1}{\sqrt{3}\kappa T_{1}\sqrt{\rho}}+\frac{1}{T_{2}^{2}\kappa^{2} \rho})=0$, then crossing may easily occur and hence  we can see that the density approaches to
 constant i.e. $\rho =\frac{3T_{1}^2}{\kappa^2 (T_{2})^4}= \frac{3\Lambda_{0}^2}{\kappa^{2}(3\zeta_{0}-\Lambda_{1})^2}$.  
 If $\bar\gamma$ is around zero, then from Eq. (56)  one can notice that the energy density $\rho$ approaches to constant i.e. $\rho
 \rightarrow \frac{3H_{0}^2}{\kappa}$ as $(t-t_{0}) \rightarrow 0$ and hence sound speed approaches to constant in the early
 time. In this way, we are able to find out the exact solutions  of the Einstein field equations in a unified manner for two different
 phases of the evolution of universe for flat FLRW model.\\

\appendix{Appendix}

\subsection*{When $T_{1} \rightarrow \infty$: Interpretation} 
In this case for {\it inflationary} phase the solution of Eq. (15) is in the form of
\begin{equation}
\label{eq58}  H^{2} = (R_{0}/R)^{2\beta_{0}}\left(H^{2}_{0} -\frac{1}{\beta_{0} T_{2}^{2}}\right) + \frac{1}{\beta_{0} T_{2}^{2}},
\end{equation}
and in term of $H(z)$ 
\begin{equation}
\label{eq59}  H^2(z) = \left(H^2_{0} -\frac{1}{\beta_{0} T^2_{2}}\right) (1+z)^{2\beta_{0}} + \frac{1}{\beta_{0} T^2_{2}}.
\end{equation}
The $H_{d}(z)$ for the dark energy is 
\begin{equation}
\label{eq60}  H^2_{d}(z) = H^2_{0}[\bar \Omega_{2} (1+z)^{2\beta_{0}} + (1-\bar \Omega_{2})],
\end{equation}
where $\bar \Omega_{2}= (1-\frac{1}{\beta_{0}H^2_{0}T^2_{2}}).$\\
It should be noted that the Eq. (60) is valid for $\bar \gamma \ne 0$, and for $\bar\gamma=0$ by directly solving
\begin{equation}
\label{eq61}  H' = -\frac{3\bar{\gamma}}{2}\frac{H}{R}+ \frac{1}{T_{2}^{2}} \frac{1}{HR},
\end{equation}
gives 
\begin{equation}
\label{eq62}  H^2_{d}(z) = H^2_{0}\left[1-\frac{2}{H^2_{0} T^2_{2}} ln(1+z)\right].
\end{equation}
The {\it H-z} relation is 
\begin{equation}
\label{eq63}  H^2 = H_{0}^2 \left\{ \Omega_{m} [\bar \Omega_{2} (1+z)^{2\beta_{0}} + (1-\bar \Omega_{2})] + (1-\Omega_{m})\right\}.
 \end{equation}
In this case if we solve the Friedmann equations with EOS $p=(\gamma-1)\rho$ without the  cosmological constant term $\Lambda$  then $\bar \Omega_{2} \rightarrow 1$ when  $T_{2}\rightarrow \infty$.  Then Eq. (63) gives
\begin{equation}
\label{eq64}  H^2_{y}(z) = H^2_{0} (1+z)^{2\beta_{0}}.
\end{equation}
For the combination  of dark matter and dark energy we can write
\begin{equation}
\label{eq65}  H^2 =   \Omega_{m} H^2_{y}(z) + (1-\Omega_{m})H^2_{0}, 
 \end{equation}
 which is exactly similar to Eq. (63). 

However, for the {\it radiation} dominated phase when $T_{1}\rightarrow \infty$ we get the analogous results.\\

\section*{Acknowledgments}
GSK is thankful to The Tata Institute of Fundamental Research (TIFR), 
Center for Applicable Mathematics, Bangalore, India  and  the Inter-University 
Center for Astronomy and Astrophysics (IUCAA), Pune, India for providing the 
necessary literature. SR is thankful to the Inter-University Centre for
Astronomy and Astrophysics (IUCAA), Pune, India as well as The Institute of
Mathematical Sciences (IMSc), Chennai, India for providing
Visiting Associateship under which a part of this work was carried
out. XHM is partly supported by the NSFC.

\end{document}